\documentclass[%
 reprint,
superscriptaddress,
 amsmath,amssymb,
 aps,
]{revtex4-2}

\usepackage{graphicx}% Include figure files
\usepackage{dcolumn}% Align table columns on decimal point
\usepackage{bm}% bold math
\usepackage{mathptmx}
\usepackage{xcolor}
\newcommand{\cor}[1]{{\color{black} #1}}

\begin{document}

\preprint{APS/123-QED}

\title{\cor{Observation of attractor transitions in active magnon-polaritons under microwatt drives}}% Force line breaks with \\

\author{Hao Wu}
 \email{hao.wu@bit.edu.cn}
 \affiliation{Center for Quantum Technology Research, Key Laboratory of Advanced Optoelectronic Quantum Architecture and Measurements (MOE), School of Physics, Beijing Institute of Technology, Beijing 100081, China}%Lines break automatically or can be forced with \\
 \affiliation{Center for Interdisciplinary Science of Optical Quantum and NEMS Integration, State Key Laboratory of Chips and Systems for Advanced Light Field Display, School of Physics, Beijing Institute of Technology, Beijing 100081, China}
 \affiliation{Beijing Key Laboratory of Quantum Matter State Control and Ultra-Precision Measurement Technology, School of Physics, Beijing Institute of Technology, Beijing 100081, China}
 
\author{Qichun Liu}%
 \affiliation{Beijing Key Laboratory of Fault-Tolerant Quantum Computing, Beijing Academy of Quantum Information Sciences, Beijing 100193, China}

\author{Yuanbin Fan}
 \affiliation{Quantum Photonics Group, Beijing Academy of Quantum Information Sciences, Beijing 100193, China}

\author{Yulong Liu}
 \email{liuyl@baqis.ac.cn}
 \affiliation{Beijing Key Laboratory of Fault-Tolerant Quantum Computing, Beijing Academy of Quantum Information Sciences, Beijing 100193, China}

\author{Qing Zhao}
 \email{qzhaoyuping@bit.edu.cn}
 \affiliation{Center for Quantum Technology Research, Key Laboratory of Advanced Optoelectronic Quantum Architecture and Measurements (MOE), School of Physics, Beijing Institute of Technology, Beijing 100081, China}
 \affiliation{Center for Interdisciplinary Science of Optical Quantum and NEMS Integration, State Key Laboratory of Chips and Systems for Advanced Light Field Display, School of Physics, Beijing Institute of Technology, Beijing 100081, China}

\date{\today}% It is always \today, today,

\begin{abstract}

\cor{Magnon-polaritons provide a room-temperature platform for investigating nonlinear cavity quantum electrodynamics in the microwave domain, but experimentally observing controlled transitions among distinct nonlinear attractors remains challenging in conventional passive systems, where strong external driving is usually required. Here we report the observation of attractor transitions in an active magnon-polariton formed by a self-oscillating microwave cavity coupled to a  yttrium iron garne (YIG) sphere. The feedback loop supplies an internal microwave drive, while Kerr frequency pulling and Suhl-mediated magnon–magnon scattering produce an enhanced effective nonlinearity. Stability analysis using experimentally calibrated parameters reveals a rich fixed-point (FP) landscape with multiple unstable-FP phases and a triple-point region. By tuning gain across these phases, we observe the first experimental evidence of explosive growth of bistability, followed by transitions to multifrequency limit cycles, comb-like/fractal spectra, and broadband chaotic dynamics at microwatt powers. Near a critical point, magnetic-field-triggered switching between nonlinear emission states produces spectral shifts up to 162 times the bare gyromagnetic response. By enabling low-power attractor transitions and attractor-switching-amplified spectral response, active magnon-polaritons open opportunities for nonlinear microwave signal generation, high-precision sensing, and neuromorphic computing.}

\end{abstract}

\maketitle

%\tableofcontents

%\begin{description}
%\item[Key claim 1] Experimentally achieved MFCs require two-tone beating; Theoretically predicted Kerr-nonlinearity-induced MFCs have not been experimentally achieved yet; Only need DC-induced microwave to power the magnon. the non-linearity is proven to be led by magnon instead of the vdP circuit due to the different results obtained with the YIG spheres placed at different locations.
%\item[Key claim 2] One current challenge may be the high threshold of microwave
%power to launch the hybridization of driving microwave and the
%intrinsic magnon frequency comb;
%\item[Key claim 3] The previously demonstrated MFCs, solitions and chaos all based on YIG thin film. There is no report using YIG spheres;
%\end{description}

\begin{figure*}[htbp!]
    \includegraphics[width=0.8\linewidth]{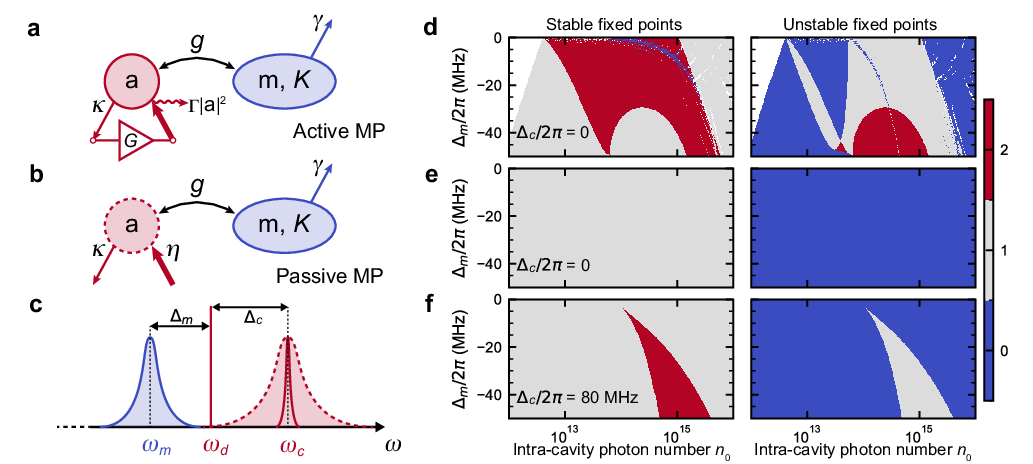}
    \caption{\textbf{Models and stability analysis of the active and passive magnon-polaritons (MP).} \textbf{a,b,} Schematics of the active and passive MP models where a cavity photon mode $a$ is coupled to a magnon mode $m$ possessing the nonlinearity $K$. The coupling is characterized by the spin-photon coupling strength $g$ and the damping of the $a$ and $m$ modes is denoted by $\kappa$ and $\gamma$ respectively. The difference between the active and passive MP arises from the driving mechanisms. For the active MP, a self-sustained $a$ mode, acting like a van-der Pol (vdP) oscillator, is established via a feedback architecture consisting of the linear gain $G$ and nonlinear gain saturation $\Gamma|a|^2$ depending on the photon number, while for the passive MP, the $a$ mode is driven externally with a strength of $\eta$. \textbf{c,} Relationship among the mode and driving frequencies. The detunings of the drive with respect to the $a$ and $m$ modes are $\Delta_c = \omega_c-\omega_d$ and $\Delta_m = \omega_m - \omega_d$, where $\omega_c$, $\omega_m$ and $\omega_d$ are the angular frequencies of the $a$ and $m$ modes, and drive. $\Delta_c = 0$ for the active MP due to the self-driving. The mode envelope of the cavity with the active (red solid curve) setting is narrower compared to the one with the passive setting (red dashed curve) due to the reduced photon damping. \textbf{d,e,f,} Simulated phase diagrams of the stable and unstable FPs in $\Delta_m-n_0$ space of the active (\textbf{d}) and passive MP (\textbf{e,f}). The color bar indicates the number of the stable/unstable FPs.}
    \label{fig:1} 
\end{figure*}

\begin{figure*}[htbp!]
    \includegraphics[width=0.8\linewidth]{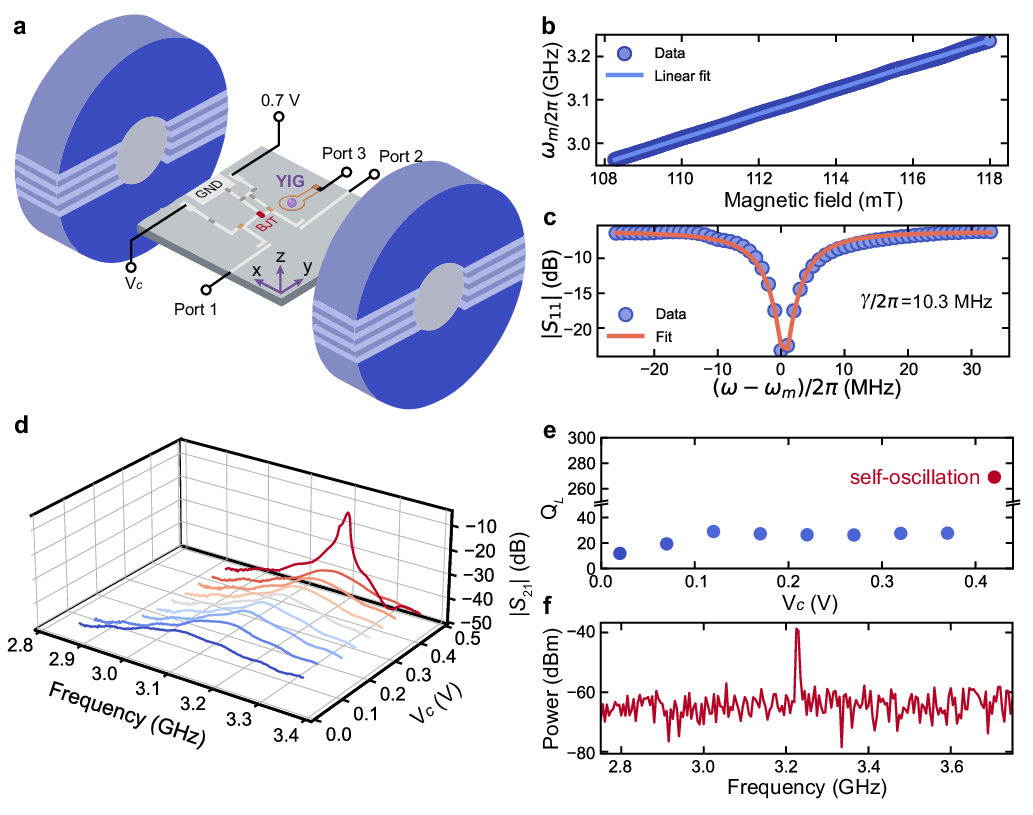}
    \caption{\textbf{Experimental setup and microwave characterizations of the magnon and bare cavity modes.} \textbf{a,} The half-wavelength microstrip-line cavity is fabricated on a printed circuit board (PCB) and the gain is provided by a bipolar junction transistor (BJT) which is powered by two-port DC voltages (one is fixed at 0.7 V and the other $V_c$ is adjustable). A yttrium iron garnet (YIG) sphere surrounded by a loop antenna is fixed in a sample holder (not shown) above the PCB and movable via a three-axis (x,y and z) translation stage. The whole setup is positioned between a pair of electromagnets supplying horizontal static magnetic fields.  There are three RF ports used for the microwave characterizations. Ports 1 and 2 are used for the microwave transmission measurements. Port 1 alone is used for recording the vdP oscillator output. Port 3 is employed for the reflection measurements of the YIG magnon mode. \textbf{b,} Linear magnetic-field-dependence of the magnon mode frequency. \textbf{c,} The damping rate of the magnon mode is obtained by fitting the microwave reflection spectrum based on a standard input-output model. \textbf{d,e,} The microwave transmission spectra and the corresponding loaded quality factors ($Q_L$) of the cavity under different $V_c$ supplied to the BJT. The drastic increase of $Q_L$ marked in red indicates the transition of the device from a passive cavity to a vdP oscillator. \textbf{f,} The microwave output spectrum of the device under the self-oscillation condition highlighted in \textbf{e}.}
    \label{fig:2} 
\end{figure*}

\begin{figure*}[htbp!]
    \includegraphics[width=0.8\linewidth]{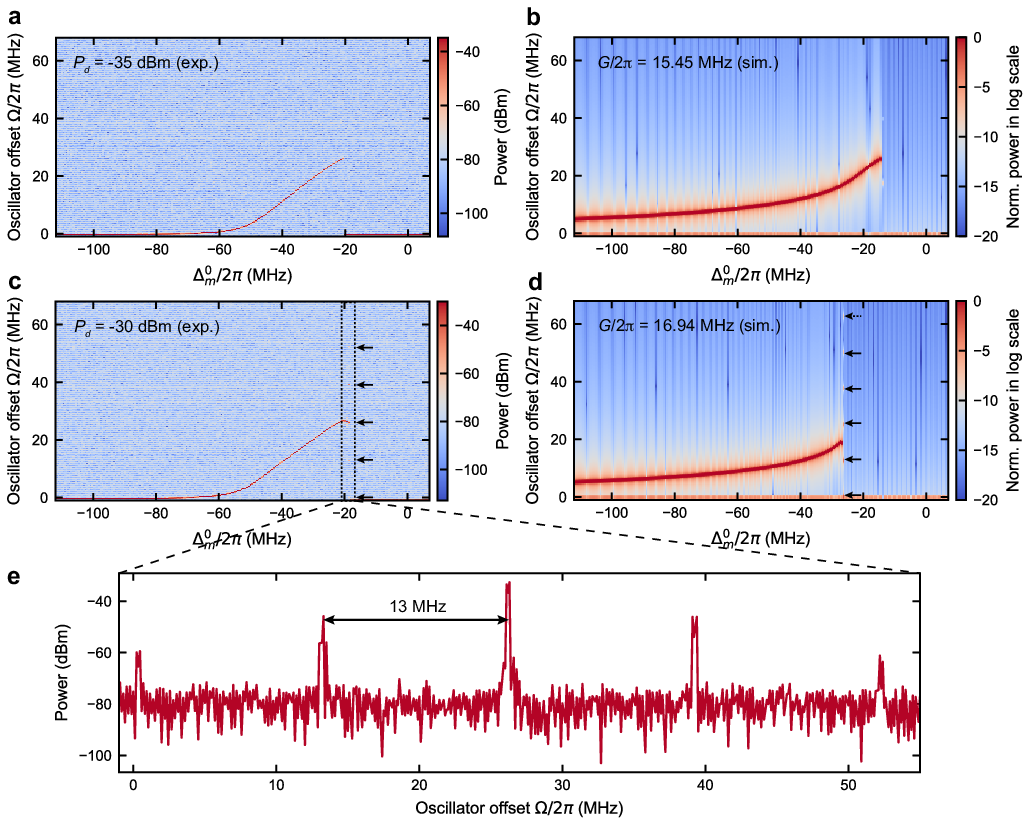}
    \caption{\textbf{\cor{Onset of attractor transitions} under microwatt-level drive.} \textbf{a,b,c,d,} Experimental and simulated results of the frequency offset of the vdP oscillator caused by its coupling to the YIG sphere subject to an up-sweeping magnetic field. $\Delta_m^0 = \omega_m - \omega_d^0$ is the magnon-drive (cavity) detuning between the magnetic-field-dependent magnon mode frequency and the initial vdP oscillator frequency. The driving power $P_d$ is varied by the gain in the circuit tuned by $V_c$ and determined according to the vdP oscillator output power. The adjustment of $P_d$ in the simulation is realized by tuning the gain factor $G$. \textbf{e,} Zoom-in of the sidebands observed in the dashed box of \textbf{c}. }
    \label{fig:3} 
\end{figure*}

\section{Introduction}
Nonlinearity is central to microwave technology, underpinning key functions such as frequency conversion, signal generation\cite{maas2003nonlinear}, and quantum-limited microwave detection\cite{clerk2010introduction}. Beyond conventional circuits built from discrete components, hybrid quantum systems\cite{clerk2020hybrid,lachance2019hybrid,awschalom2021quantum} offer new opportunities to tailor nonlinear response and microwave functionalities \textit{in situ} via external field tuning\cite{khan2021coupling} and cavity mode engineering\cite{goryachev2014high,PhysRevLett.113.083603}. The magnon-polaritons (MP), formed by the strong coupling of collective spin waves in YIG to microwave photons, provide one such room-temperature platform where nonlinearities normally originate from the spin system. One mechanism is the Suhl instability\cite{suhl1957theory}, in which sufficiently strong pumping of the uniform precession triggers magnon–magnon scattering into secondary spin-wave modes; in typical bulk geometries these thresholds are often milliwatt (mW)-class\cite{bryant1988spin,wiese1990multistability}, and the experiments were generally not designed to form well-resolved, strongly coupled magnon–photon hybrids. Moving into the strong-coupling regime, the nonlinear signatures of MP arising from the Suhl instability have started to be explored recently, but so far only the drive-induced suppression of the normal-mode splitting has been observed\cite{lee2023nonlinear,wagle2026deeply}. A second and widely discussed nonlinear mechanism is the magnon Kerr effect giving rise to driving-amplitude-dependent frequency shift of the magnon mode\cite{zhang2019theory}. Experimentally, Kerr-driven MP nonlinearity has been observed most robustly as bistability\cite{wang2018bistability,pan2022bistability}, whereas theoretically anticipated regimes such as frequency combs\cite{liu2018magnon} and chaos\cite{liu2019phase} remain experimentally elusive if without additional coupling to other quasiparticles, like phonons\cite{xu2025kerr,peng2024cavity}. This gap is especially evident for spherical YIG samples whose macroscopic mode volume and intrinsically small magnetocrystalline anisotropy yield small Kerr coefficients\cite{xu2023magnonic}, pushing the nonlinear thresholds upward. %In this context, thin-film approaches that enhance effective nonlinearity are emerging\cite{petrosyan2026magnon}, but the exploration of rich nonlinear dynamical phases is still at an early stage.

Both the Suhl- and Kerr-mediated routes to nonlinear dynamics in conventional (passive) MP systems tend to rely on power-hungry external drives, indicating that photon storage accomplished by cavities is still insufficient to support intense driving fields for accessing broad nonlinear landscapes. This motivates making the hybrid system active\cite{yao2017cooperative} by supplying gain to the cavity mode for mitigating the photon damping and thus boosting the number of microwave photons coupled to magnons. Active MP platforms have already enabled striking capabilities, such as coherent emission\cite{yao2023coherent,wang2025single} and nonreciprocal amplification\cite{wang2023realization}, yet these demonstrations have remained in the regime where the intrinsic magnon nonlinearities are neglected. Achieving and controlling richer nonlinear \cor{attractors} in active MP\cite{wang2019magnetic} therefore remains an open experimental goal.

Here we overcome these constraints by activating YIG-sphere-based MP in a chip-scale direct-current (DC)-powered feedback architecture operated above the self-oscillation threshold, so that the efficient drive is realized internally without an external microwave generator. Meanwhile, the sustained oscillation field coupled to the YIG magnon mode also acts as an \textit{in situ} probe to reveal the MP status under magnetic-field tuning through its emission properties. By combining this self-oscillation drive with operation in a bias-field regime that favours the Suhl instability, while retaining Kerr-induced frequency pulling, we access \cor{deeply} nonlinear MP dynamics at microwatt ($\mu$W)-level injected powers. We observe the full route of the MP nonlinear dynamics to chaos, together with an explosive growth of bistability and electrically tunable \cor{attractor} transitions via DC-controlled gain and detuning. This work establishes active MP as a compact, low-power platform for exploring complex dynamics in the microwave domain at ambient conditions and lays the foundation of the active-MP-based nonlinear microwave devices.

%\subsection*{Model and stability analysis}

\section{Model and stability analysis} 
We first theoretically verify the ease of the active MP to reveal nonlinear dynamics compared with the passive system. Figs.~\ref{fig:1}a and \ref{fig:1}b show the models of the active and passive MP, where both systems include a cavity photon mode $a$ coupled to a magnon mode $m$ (e.g., the Kittle mode of YIG spheres)  with a spin-photon coupling strength $g$. Here, we consider the magnon mode possessing an intrinsic nonlinearity $K$, for instance the Kerr nonlinearity\cite{zhang2019theory}, and the damping rate $\gamma$. What differs between the two systems is the photon dynamics, for which the photon damping $\kappa$ in the active MP is compensated by the gain $G$ and the amplified photon flux is fed back to the cavity. To avoid the infinite growth of the photon occupation in the cavity, a gain saturation term $\Gamma|a|^2$ is also introduced so that the equation of motion describing the photon mode (without the coupling to the magnon mode) resembles the classical van-der Pol (vdP) oscillator model\cite{van1926lxxxviii} (see \textbf{Appendix~\ref{appendix_eom}}). In contrast, the photon mode in the passive system undergoes the damping controlled by the cavity and an external driving term $\eta$ is employed to build up the intra-cavity field and initialize the nonlinear dynamics. Another difference between the two systems is the detuning settings shown in Fig.~\ref{fig:1}c. For the active MP, as the drive can be provided by the cavity mode itself once the gain is sufficient to activate the self-oscillation, the driving frequency $\omega_d$ always matches the cavity mode frequency $\omega_c$, resulting the detuning $\Delta_c = \omega_c-\omega_d= 0$, while $\Delta_c$ is an arbitrary value depending on the choice of the external drive for the passive MP. The complete equations of motion describing the active and passive MP with the parameters settings can be found in \textbf{Appendices~\ref{appendix_eom} and \ref{appendix_ss}}.

Based on the models, we conduct the stability analysis\cite{khan2018frequency} around the FPs of the two systems to evaluate the nonlinear behaviors; details can be found in \textbf{Appendix~\ref{appendix_ss}}. \cor{The fixed-point analysis identifies where stationary MP states remain stable and where they become susceptible to transitions into time-dependent attractors.} As the magnon-associated nonlinearity depends on the driving field, for the fair comparison, we plot the phase diagrams of the stable and unstable FPs for both systems in $\Delta_m-n_0$ space (Figs.~\ref{fig:1}d-f), where $\Delta_m = \omega_m-\omega_d$ is the detuning between the magnon mode and the drive and $n_0$ is the intra-cavity photon number which is associated with the feedback loop setting in the active system and the external drive in the passive system. We find that under the resonant drive ($\Delta_c=0$), the active MP reveals multiple regimes in both phase diagrams of the stable and unstable FPs (Fig.~\ref{fig:1}d), while only a stable FP of the passive MP emerges (Fig.~\ref{fig:1}e) indicating the system sits in the purely linear regime. Although the increase of the detuning $\Delta_c/2\pi$ to 80 MHz gives rise to a new phase that endows the passive MP bistability (Fig.~\ref{fig:1}f), i.e., two stable FPs and one unstable FP, its bistable regime is much smaller and the driving threshold reflected by $n_0>10^{14}$ is more than one order of magnitude higher compared to those in the active MP. Moreover, the active MP can completely lose stable FPs or admit one stable FP coexisting with two unstable FPs. Those phases would facilitate transitions of the active MP dynamics to time-dependent attractors that enable nonlinear phenomena beyond bistability\cite{pastor1993ordered}.

%\section*{Experimental setup and characterizations}

\section{Experimental setup and characterizations} 
The experimental setup for constructing the active MP is shown in Fig.~\ref{fig:2}a, where a half-wavelength microstrip-line cavity with a gain element, i.e., a bipolar junction transistor (BJT) is fabricated and a 1-mm-diameter YIG sphere is placed above the cavity whose position can be finely adjusted to achieve the optimal coupling between the photon and magnon modes; details in \textbf{Appendix~\ref{appendix_exp}}. The magnon mode frequency $\omega_m$ can be tuned by an electromagnet and obtained with the reflection measurements. By linear fitting the magnetic-field-dependent $\omega_m$ shown in Fig.~\ref{fig:2}b, the magnon mode follows a Kittle formula $\omega_m = \gamma_e\mu_0(H_0+H_A)$, where $H_0$ is the applied static magnetic field, $\gamma_e/2\pi = 28.2$ MHz/mT is the gyromagnetic ratio and $\mu_0H_A = -3.35$ mT is the magnetocrystalline anisotropy filed. Fig.~\ref{fig:2}c shows the reflection spectrum can also be used to obtain the magnon damping rate $\gamma/2\pi=10.3$ MHz by fitting it based on the input-output theory (see \textbf{Appendix~\ref{appendix_exp}}). 

In addition to the magnon mode properties, we also conduct the microwave characterizations of the bare cavity mode (without coupling to the magnon mode). The cavity transmission spectra (Fig.~\ref{fig:2}d) obtained with the different voltages $V_c$ supplied to the BJT show the cavity mode frequency $\omega_c/2\pi$ ranges from 3.1 to 3.2 GHz and the $V_c$-controlled gain of the BJT can reduce the insertion loss and improve the loaded quality factor $Q_L$ of the cavity. Fig.~\ref{fig:2}e shows $Q_L$ can be increased by a factor of three (10 to 30) when $V_c$ is increased close to 0.4 V. Once $V_c$ is above 0.4 V, a drastic increase of $Q_L$ to 275 is observed accompanied with a surge of the transmission amplitude ($|S_{21}|$) shown in Fig.~\ref{fig:2}d. The sudden change of the transmission property is due to the self-oscillation of the cavity when the gain is sufficient to overcome the photon damping in the circuit. Without connecting to any external microwave sources, we observe a microwave emission peak with a power of -40 dBm around 3.2 GHz as shown in Fig.~\ref{fig:2}f, which implies the cavity is configured to be a vdP oscillator. Our measurements of the active MP discussed in the following context are implemented with the vdP oscillator, by which the emission can be employed to drive the YIG sphere and simultaneously behaves as the probe for evaluating the MP dynamics as the information of the magnon mode, e.g., the magnon frequency is coded in the emission spectra once the MP is formed. 

%\section*{Active MP under microwatt-level drive}

\section{Nonlinear active MP under microwatt-level drive}\label{weak drive}
It is found that the position of the YIG sphere with respect to the vdP oscillator is crucial to realize the strong spin-photon and thus form the MP (see \textbf{Supplemental Material}). When the optimized position is determined, we first investigate the nonlinear behaviors of the active MP under weak drives that no larger than -30 dBm, i.e., 1 $\mu$W. As shown in Fig.~\ref{fig:3}a, under a -35-dBm drive, the vdP oscillator frequency undergoes a blue shift $\Omega/2\pi$ up to 26 MHz before an abrupt switch back to the initial condition when the magnetic field is swept up continuously. This result indicates the formation of MP\cite{yao2023coherent}, for which the driving power needed to achieve the nonlinear characteristics, such as the memory effect and bistability revealed by the up- and down-sweeping results (see \textbf{Supplemental Material}), is reduced by about four orders of magnitude compared to the previously reported active MP\cite{zhang2024van}. However, the switching point obtained here locates at a negative magnon-drive (cavity) detuning $\Delta_m^0/2\pi = -20$ MHz that differs from the reported positive ones\cite{yao2023coherent,zhang2024van}. 

Remarkably, when the drive is increased further to -30 dBm by increasing $V_c$, nonlinear features that were not observed from the active MP before emerge: the blue-shifted vdP oscillator frequency starts to bend and sidebands appear when approaching the switching point, shown in Figs.~\ref{fig:3}c and \ref{fig:3}e. The bending behavior is analogous to the Kerr-induce magnon frequency shift found in the passive MP system with typically 25-dBm drives\cite{wu2022observation,wang2018bistability}. The appearance of the sidebands with a frequency spacing of 13 MHz clearly shows the active MP lose stability and enters into the regime that the attractor of the multifrequency limit cycle exists\cite{khan2018frequency}. We can exclude the contribution of the vdP oscillator's nonlinearity to the observed dynamics as the single-tone emission from the magnon-decoupled oscillator is magnetic-field insensitive and does not evolves to complex patterns when $V_c$ is adjusted.

To understand the nonlinear features, we perform the simulation of the active MP dynamics (see \textbf{Appendix~\ref{appendix_sim}}) and obtain the dependence of the oscillator offset $\Omega$ on $\Delta_m^0$ with the memory setting same as in the experiments (Figs.~\ref{fig:3}b and \ref{fig:3}d). The dependence under different driving powers is simulated by tuning the gain factor $G$. We find the results with $G/2\pi = 15.45$ and 16.94 MHz qualitatively reproduce Figs.~\ref{fig:3}a and \ref{fig:3}c, where the switching points at negative detunings, frequency bending and sidebands are all observed. In particularly, the simulated frequency spacing of the sidebands agrees well with the experiment except the marginal sideband is missing in the experiment which might arise from its low signal-to-noise ratio. According to the simulation, we obtain the gain saturation factor $\Gamma/2\pi=0.8~\mu$Hz and the spin-photon coupling strength $g/2\pi = 25$ MHz which surpasses the photon and magnon damping indicating the MP settle into the strong coupling regime. It is worth noting that the 'Kerr' nonlinearity coefficient $K/2\pi = 3.2~\mu$Hz extracted from the simulation is unexpectedly large for YIG spheres which is normally of the order of nHz\cite{petrosyan2026magnon}. We attribute it to additional material-based nonlinear mechanisms involved. The most possible ones are the magnon-magnon scattering processes\cite{zheng2023tutorial} as the corresponding Suhl instability occurs when the system operates below the critical frequency of 3.27 GHz for YIG spheres\cite{wagle2026deeply}. Our MP prepared within the range from 3.1 to 3.2 GHz fulfill the criteria. Moreover, the coexistence of the self-Kerr effect and the Suhl instability for promoting the complex nonlinear dynamics of YIG has been theoretically validated\cite{elyasi2020resources}. Thus, herein we assign 3.2 $\mu$Hz to be the effective nonlinearity coefficient of our active MP system.

%With all system parameters determined, we further conduct the stability analysis of the active MP in real conditions. The phase diagrams of the stable and unstable FPs in $\Delta_m-G$ space (Fig.~\ref{fig:3}f) reveal that increasing the gain would expand the bistable regime especially for the large magnon-drive detuning ($\Delta_m <-45$ MHz), while facilitating the instability implied by the increased number of the unstable FPs up to three. We find the triple-point phase transition of the unstable FPs occurring at $G/2\pi = 16.67$ MHz and $\Delta_m/2\pi = -45$ MHz.%, which are quite close to the parameter settings obtained with the -30-dBm drive that lead to the MP sidebands. 
%Note that, in Figs.~\ref{fig:3}c and \ref{fig:3}d, $\Delta_m^0$ refers to the detuning between the magnon mode and the \textit{initial} oscillator frequency, thus the critical detuning in real time $\Delta_m = \Delta_m^0 - \Omega$ is approximately -45 MHz. \cor{Hence, the MP sideband onset at $P_d$= -30 dBm, corresponding to $G/2\pi = 16.94$ MHz, links the multifrequency response to the crossing of unstable-FP phases leading to the transition from the bistable to multifrequency-limit-cycle attractor.}

%\section*{Route to chaotic active MP}

\begin{figure*}[htbp!]
    \includegraphics[width=0.8\linewidth]{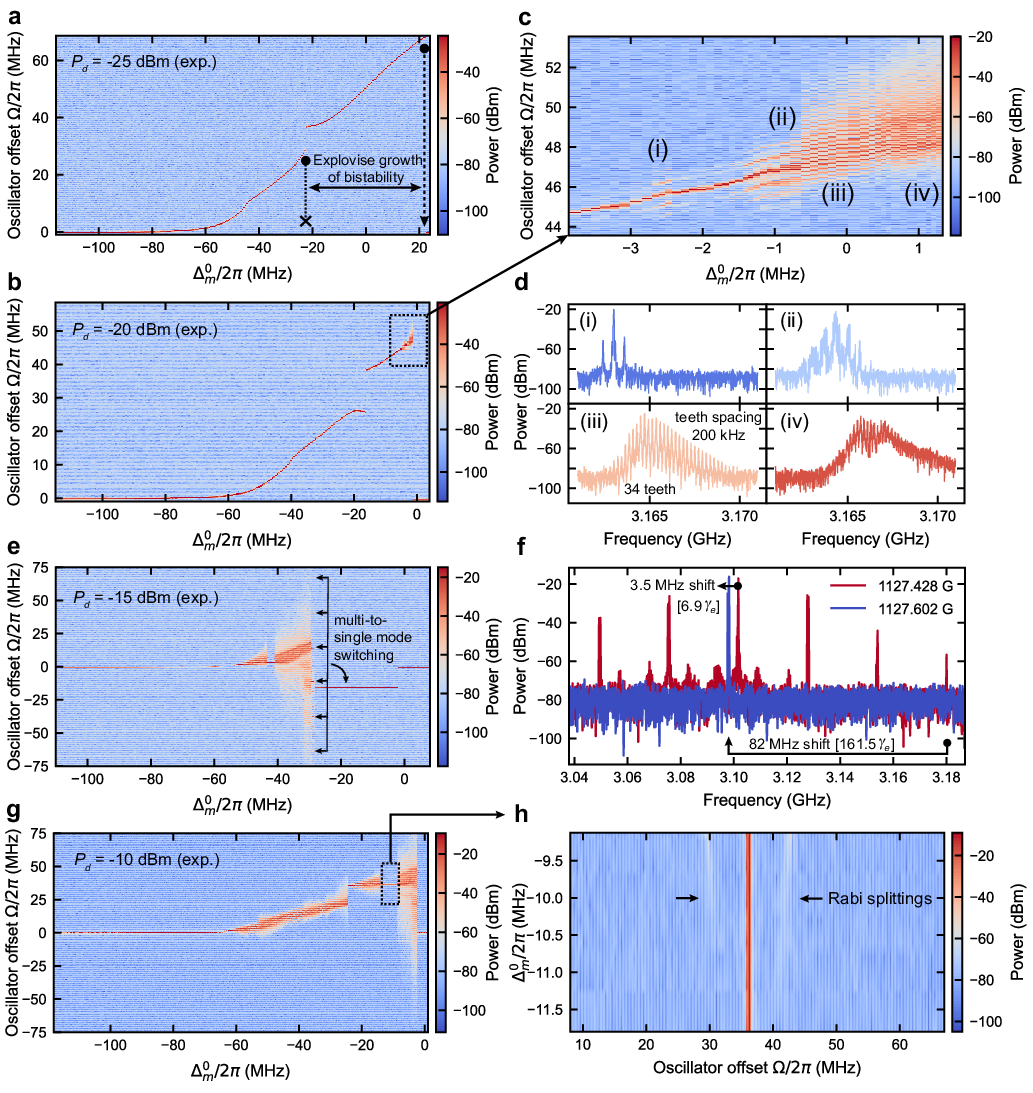}
    \caption{\textbf{Route to chaotic active MP.} \textbf{a,b,e,g,} Experimental results of the frequency offset of the vdP oscillator coupled to the YIG sphere subject to a up-sweeping magnetic field under the driving powers beyond 1 $\mu$W (-30 dBm). \textbf{c,d,} Zoom-in of the rich nonlinear phenomena observed in the dashed box of \textbf{b} where the four nonlinear regimes show (i) multifrequency limit cycles, (ii) fractals, (iii) frequency-comb-like structure and (iv) chaos. \textbf{f,} Switching of the vdP output spectrum from the multimode to single-mode emission observed in \textbf{e}. The significant peak shifts occur within a change of the magnetic field less than 0.2 G. \textbf{h,} Zoom-in of the magnetic-field-dependent Rabi splittings of the vdP oscillation observed in \textbf{g}. }
    \label{fig:4} 
\end{figure*}

\section{Route to chaotic active MP} 
We further explore the deeply unstable regime predicted by the stability analysis with the driving power higher than -30 dBm and for the first time observe a complete route to chaotic active MP, as shown in Fig.~\ref{fig:4}. Under a -25-dBm drive, we observe a jump of the oscillator frequency to a even higher state at $\Delta_m^0\sim-20$ MHz (Fig.~\ref{fig:4}a) instead of returning to the initial state as demonstrated in Fig.~\ref{fig:3}a and \ref{fig:3}c. The blue shift of the emission would continue when $\Delta_m^0$ passes zero to the positive regime and eventually get back to the initial condition at $\Delta_m^0=+21$ MHz. The whole process results in a record shift, $\Omega/2\pi= 68$ MHz, of the active MP state\cite{yao2023coherent,zhang2024van} and this intriguing phenomenon is so-called the explosive growth of bistability which has only been theoretically anticipated requiring a driving power more than 500 mW (27 dBm) \cite{bi2025explosive}. The explosive growth of the bistable regime is confirmed by comparing the up- and down-sweeping results (see \textbf{Supplemental Material}).

As the driving power is increased to -20 dBm, in addition to the Kerr-induced magnon frequency bending and the explosive growth of bistability, complex patterns appear near the switching point at the detuning $\Delta_m^0=-1$ MHz shown in Fig.~\ref{fig:4}b. The details of the patterns emerge when we finely scan the magnetic field and four different nonlinear regimes are progressively revealed with $\Delta_m^0/2\pi$ swept
from -3 to above 1 MHz (Fig.~\ref{fig:4}c). The discrepancy of $\Delta_m^0$ showing the patterns in Figs.~\ref{fig:4}b and \ref{fig:4}c is due to the different intervals chosen for the magnetic-field sweep, demonstrating another crucial nonlinear feature that the evolution of the MP dynamics is \textit{path dependent}. Fig.~\ref{fig:4}d clearly shows \cor{a detuning-controlled progression across different nonlinear attractors}, where the MP dynamics undergoes a four-stage transition: (i) multifrequency limit cycle; (ii) fractals; (iii) frequency-comb-like structure and (iv) chaos. The comb-like structure includes 34 teeth with a uniform spacing of 200 kHz.

Further strengthening the drives to -15 dBm and -10 dBm gives rise to a broader range of the complex patterns (Figs.~\ref{fig:4}e-h), especially chaos across $\Delta_m^0$. For both drives, the onsets of the nonlinear MP dynamics losing stability occur at $\Delta_m^0/2\pi$ around -60 MHz that agrees well with the detuning of the deeply unstable regime shown in Fig.~\ref{fig:3}f. For the -15-dBm drive, we find the chaotic emission spectra can span up to 150 MHz (Fig.~\ref{fig:4}e) and then evolve to multiple emission peaks with a spacing of 28 MHz at $\Delta_m^0/2\pi = -29$ MHz before a sudden transition to a single peak within a magnetic-field change of less than 0.2 G (Fig.~\ref{fig:4}f). There are two interesting properties of the phase transition within such a narrow field range. One is the single emission peak comes from a newly emerged MP state with a 15-MHz \textit{red} shift respect to the original state where the magnon and photon modes are decoupled. It stays magnetically \textit{insensitive} across a range of $\Delta_m^0$ around 26 MHz before getting back to the original state. Conversely, during the multi-to-single emission mode switching, the MP manifests an ultrahigh magnetic sensitivity that the main emission peak and the marginal sideband experience 3.5 and 82 MHz shifts, which correspond to 6.9 and 161.5 times of the gyromagnetic ratio ($\gamma_e$), respectively. 

Under the -10-dBm drive, the landscape of the MP dynamics at $\Delta_m^0/2\pi>-60$ MHz  is almost chaotic (Fig.~\ref{fig:4}g) except for $\Delta_m^0/2\pi$ between -11.5 and -9 MHz where the magnetic-field-dependent Rabi oscillations are observed forming the MP triplet (Fig.~\ref{fig:4}h). The MP triplet\cite{yao2017cooperative} indicates the magnon-photon cooperativity enhancement offered by the active architecture.

\begin{figure*}[tp!]
    \includegraphics[width=1.0\linewidth]{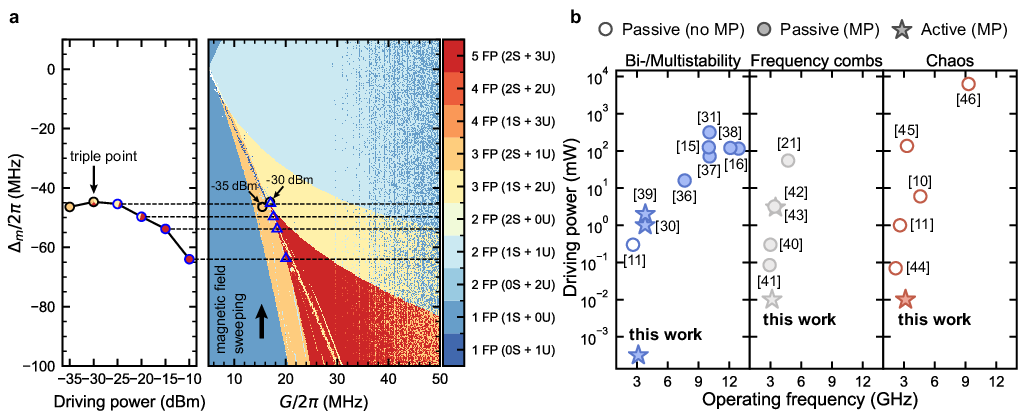}
    \caption{\cor{\textbf{Power-dependent stability phase diagram of the active MP revealing its power efficiency in achieving nonlinear attractor transitions.} \textbf{a,} Dependence of the critical detuning in real time $\Delta_m = \Delta_m^0 - \Omega$ measured at different driving powers (left panel). Simulated phase diagrams of the stable and unstable FPs in $\Delta_m-G$ space of the active MP (right panel). The parameters used for generating the phase diagrams are obtained from the simulation shown in Figs.~\ref{fig:3}{b} and \ref{fig:3}{d} that reproduce the experimental results. The color bar consistent with the circle fillings in the left panel indicates the different FP situations where S and U stand for stable and unstable FPs respectively. The deterministic phases of the active MP under -30 and -35-dBm drives at the critical detunings are labeled with circles. Blue triangles emulate the shift of the critical detuning, where the '2S+1U' to '2S+3U' phase transition emerges, as the gain/driving power is increased.} \textbf{b,} Summary of the experimental studies demonstrating microwave nonlinear phenomena observed from YIG spheres. Those studies are compared in terms of the microwave driving power and operating frequency required to observe the three nonlinear phenomena including the bi-/multistability\cite{wang2018bistability,pan2022bistability,shen2022mechanical,shen2021long,zhang2024van,hyde2018direct,rao2023meterscale,wu2022observation,wiese1990multistability}, frequency combs\cite{wang2024enhancement,xu2023magnonic,zhao2025experimental,rao2023unveiling,gui2025broadband} and chaos\cite{bryant1988spin,wiese1990multistability,carroll1989chaos,mitsudo1992period,rodelsperger1995route}. The systems based on the passive and active architectures are represented by circles and stars respectively, among which the passive systems with and without forming the MP are distinguished by the filling condition of the circles.}
    \label{fig:5} 
\end{figure*}

\cor{\section{Power-dependent nonlinear attractor transitions}}

\cor{The distinct nonlinear dynamics observed through Fig.~\ref{fig:3} to Fig.~\ref{fig:4} implies the power-dependent nonlinear attractor transitions in the active MP. In Fig.~\ref{fig:5}a, we show the critical detuning in real time $\Delta_m = \Delta_m^0 - \Omega$, that induces the abrupt switching of the single emission tone or the evolution from the single to multiple/chaotic emission tones, shifts to the deeper negative regime as the driving power is increased beyond -30 dBm. Under -10-dBm drive, the single-frequency limit-cycle attractor of the MP transforms to the chaotic attractor at a maximum negative critical tuning of -70 MHz which places the magnon-photon coupling into the dispersive regime where $|\Delta_m|\gg |g|$\cite{blais2004cavity}.}

\cor{With all system parameters determined in Section.~\ref{weak drive}, we further conduct the stability analysis of the active MP to understand the power-dependent nonlinear attractor transitions. The phase diagrams of the stable and unstable FPs in $\Delta_m-G$ space (right panel of Fig.~\ref{fig:5}a) reveal that increasing the gain (i.e., the driving power) would convert the conventional bistable phase comprising two stable FPs (2S) and one unstable FP (1U) into the phases '1S+2U' and '2S+3U' possessing the enhanced instabilities where the unstable FPs can be increased up to three. This phase transition agrees well with the occurrence of the multiple-attractor-induced nonlinear dynamics at elevated driving powers. 

Moreover, for the nonlinear MP dynamics obtained under -35 and -30 dBm shown in Fig.~\ref{fig:3}, the fitted gain factors  $G/2\pi = 15.45$ and 16.94 MHz and the measured critical detuning $\Delta_m/2\pi=-46.4$ and -44.7 MHz indicate the two power settings drive the MP dynamics into different stability phases. The active MP enter into the conventional bistable phase ('2S+1U') with -35-dBm drive but settle into the triplet point, i.e., the boundary among the phases '2S+1U', '1S+2U' and '2S+3U' under -30-dBm drive where, in additional to the bistable one, the onset of the nonlinear attractor of multifrequency limit cycle is observed. The further increase of the gain beyond the triplet point would facilitate the '2S+1U' to '2S+3U' phase transition appearing at the territory with the enlarged negative critical detuning $\Delta_m$ that also matches the observed power-dependence of $\Delta_m$ shown in the left panel of Fig.~\ref{fig:5}. Note that, the slightly enhanced drive power beyond -30 dBm, e.g., the -25 dBm would drive the MP from the bistable phase into a noisy phase regime where besides the continuous '1S+2U' and  '2S+3U' phases, the finely fragmented phases co-exist. This may explain no strange attractors are observed in Fig.~\ref{fig:4}a, instead, the explosive growth of bistability accompanied with the MP emission frequency jump is monitored, which is extremely sensitive to the environmental perturbance, e.g., the transient power disturbance caused by the switching on/off of the neighboring instrument, indicating the nondeterministic phase transitions. Increasing the driving power to -20 dBm would largely get rid of the phase uncertainty by entering the phases with more clear phase boundaries and expanded '2S+3U' regime, which promotes the unambiguous and progressive transitions from the bistable to chaotic attractors shown in Fig.~\ref{fig:4}c.}

%\section*{Discussion and outlook}

\section{Discussion and outlook} 
In summary, we have experimentally achieved deeply nonlinear active MP based on YIG spheres and \cor{observed fruitful attractor transitions} with the driving power no more than 100 $\mu$W, which is to date the lowest among the experimental studies on the nonlinear MP with YIG spheres, as summarized in Fig.~\ref{fig:5}. \cor{This addresses a key challenge in nonlinear cavity magnonics: transforming intrinsic magnon nonlinearities into a controlled sequence of nonlinear polariton states without strong external microwave driving.} Notably, owing to the chip-based feedback architecture, the microwave drives relying on the self-oscillation of the cavity mode only require small DC voltages ($<$3 V) to operate. The elimination of the bulk external microwave generators lays the foundation to achieve more compact MP systems. The mechanism leading to the rich nonlinearities of the active MP arises from the Kerr effect and the Suhl instability collectively. We realize that similar to this work, most previous investigations reporting the complex nonlinear dynamics of YIG spheres, especially chaos, lie in the frequency regime around 3 GHz (Fig.~\ref{fig:5}) where the Suhl instability is promoted. The increase of the frequency would significantly level up the driving power threshold of the nonlinearities. The active approach, complementary to the exceptional-point enhancement\cite{wang2024enhancement} and phase modulation\cite{peng2024cavity}, would offer a microwave-generator-free solution to overcome the challenge at elevated frequencies.

\cor{More broadly, our work fits into the rapidly emerging direction of self-oscillating hybrid spin–cavity systems for low-power nonlinear microwave functionalities. During the completion of this work, we are aware of the recent studies have shown that bistable-transition-point physics can enable exceptional sensitivity and ultralow-power microwave frequency-comb generation in a nitrogen-vacancy-based spin-oscillator system\cite{wang2026exceptional,wang2026ultralow}. In this context, our work extends this direction to MP, where the intrinsic material nonlinearities enable access to a substantially deeper nonlinear landscape on chip.}

Furthermore, the observation of the intermediate MP state that insensitive to magnetic fields raises interesting questions that call for future investigations. A model considering the multimode networks\cite{khan2021non} formed by the coupling between a magnon mode and several photon modes may help clarify the birth of the intermediate state right after the vanishing of multiple emission peaks. Enriching the theoretical framework of the nonlinear active MP would make it a traceable platform for studying the microwave nonlinear dynamics and benefit the design of MP-based sensing\cite{mi2025ultrasensitive} and computing devices\cite{shen2021long,chumak2022advances}. Further, it will be of great interest to explore the quantumness of the active MP, e.g., the proposed squeezing\cite{kani2025squeezed} and entanglement\cite{elyasi2020resources}.

\begin{acknowledgments}
We thank Sheng-Wen Li, Dazhi Xu, Caihua Wan and Bo Zhang for insightful discussions. We acknowledge the support of the Science Challenge Project (Grant No. TZ2025017). H.~W.~acknowledges the support of the National Natural Science Foundation of China (No.~12574382 and No.~12204040), the Young Elite Scientists Sponsorship Program of the Beijing High Innovation Plan (Grant No.~20250800), Fundamental and Interdisciplinary Disciplines Breakthrough Plan of the Ministry of Education of China (JYB2025XDXM115), the Fundamental Research Funds for the Central Universities and the Beijing Institute of Technology Research Fund Program for Young Scholars (Grant No. XSQD-6120230016). Y.~L.~acknowledges the support of the National Natural Science Foundation of China (No.~12374325 and No.~92365210), Beijing Natural Science Foundation (Z240007), and the Young Elite Scientists Sponsorship Program by CAST (Grant No.~2023QNRC001).

\end{acknowledgments}

\section*{Author Contributions}

H.~W. conducted the measurements and developed the theory. Y.~F and Q.~L fabricated and optimized the chip-based active cavity. H.~W. and Y.~L. conceived the project. Y.~L. and Q.~Z. supervised the project. All authors contributed to the interpretation of the data and reviewed the paper.

\appendix

%\section{Methods}

\section{Equations of motion}\label{appendix_eom}
We describe MP at the semiclassical mean-field level using complex mode amplitudes for a single photon mode $a(t)$ and the magnon mode $m(t)$. Unless stated otherwise, all parameters are expressed in angular-frequency units (rad\,s$^{-1}$).

%\paragraph*{Passive MP system.}
\subsection{Passive MP system}
In the frame rotating at the coherent drive frequency $\omega_d$, the passive system obeys
\begin{align}
\dot a &= -\Big(\frac{\kappa}{2}+ i\Delta_c\Big)a - i g\,m + \eta,
\label{eq:passive_a}\\
\dot m &= -\Big(\frac{\gamma}{2}+ i\Delta_m\Big)m - i g\,a - i K|m|^2 m ,
\label{eq:passive_m}
\end{align}
The drive amplitude is implemented via input--output theory as $\eta=\sqrt{\kappa_{\mathrm{ext}}}\,s_{\mathrm{in}}$, with incident photon flux $s_{\mathrm{in}}=\sqrt{P_{\mathrm{in}}/(\hbar\omega_d)}$ and external coupling rate $\kappa_{\mathrm{ext}}$ (equal to $\kappa/2$ under the critical coupling). $P_\textrm{in}$ is the input microwave power.

%\paragraph*{Active MP system.}
\subsection{Active MP system}
For the active system , in the reference frame of the bare photon oscillation (i.e., the autonomous vdP oscillation), the equation of the magnon mode is the same as Eq.~(\ref{eq:passive_m}) and the photon mode is changed to be
\begin{align}
\dot a &= \Big(G-\frac{\kappa}{2}\Big)a - \Gamma|a|^2 a - i g\,m.
\label{eq:active_a}
%\dot m &= -\Big(\frac{\gamma}{2}+ i\Delta_m\Big)m - i g\,a - i K|m|^2 m .
%\label{eq:active_m}
\end{align}
%Here $\mu$ is the linear gain rate, $\Gamma>0$ is the gain-saturation (nonlinear damping) coefficient, and $\Delta_m\equiv \omega_m-\omega_a$ is the magnon detuning from the bare photon-mode frequency $\omega_a$ in this rotating frame. In the phase-diagram calculations we consider the autonomous active MP and set $\eta=0$.

\section{Steady states and stability analysis}\label{appendix_ss}

%\paragraph*{Steady states of the passive MP.}
\subsection{Steady states of the passive MP}
Setting $\dot a=\dot m=0$ in Eqs.~(\ref{eq:passive_a})--(\ref{eq:passive_m}) yields
\begin{align}
a &= \frac{\eta}{\frac{\kappa}{2}+i\Delta_c+\frac{g^2}{\frac{\gamma}{2}+i(\Delta_m+K|m|^2)}},
\qquad
m = -\frac{i g\,a}{\frac{\gamma}{2}+i(\Delta_m+K|m|^2)}.
\label{eq:passive_ss}
\end{align}
Defining the magnon occupation $n_m\equiv |m|^2$, taking the modulus of Eq.~(\ref{eq:passive_ss}) reduces the steady-state condition to a real cubic polynomial in $n_m$. We solve this cubic and retain only real, positive roots. For each admissible $n_m$, the corresponding $(a,m)$ is reconstructed using Eq.~(\ref{eq:passive_ss}) and subsequently tested for linear stability.

To compare different cavity detunings on a common horizontal axis, we parameterize the drive strength by the ``empty-condition photons'' $n_0$, defined as the photon number of the uncoupled cavity ($g=0$):
\begin{align}
n_0 \equiv |a|^2\big|_{g=0}=\frac{|\eta|^2}{(\kappa/2)^2+\Delta_c^2}
=\frac{\kappa_{\mathrm{ext}}}{(\kappa/2)^2+\Delta_c^2}\frac{P_{\mathrm{in}}}{\hbar\omega_d}.
\label{eq:n0_passive}
\end{align}
In the simulations, Eq.~(\ref{eq:n0_passive}) is inverted to map a chosen $n_0$ to $P_{\mathrm{in}}$ for each $\Delta_c$.

%\paragraph*{Steady states of the active MP.}
\subsection{Steady states of the active MP}
For an autonomous vdP oscillator, the relevant steady behavior is a single-frequency solution. We therefore use the ansatz
\begin{align}
a(t)=a_0\,e^{-i\Omega t},\qquad m(t)=m_0\,e^{-i\Omega t},
\label{eq:ansatz}
\end{align}
with constant complex amplitudes $(a_0,m_0)$ and a real frequency offset $\Omega$ relative to the reference frame. %Substituting Eq.~(\ref{eq:ansatz}) into Eqs.~(\ref{eq:passive_m})--(\ref{eq:active_a}) yields coupled equations for $(a_0,m_0,\Omega)$.
Substituting Eq.~(\ref{eq:ansatz}) into Eqs.~(\ref{eq:passive_m})--(\ref{eq:active_a}), we obtain
\begin{align}
\Big(G-\frac{\kappa}{2}-\Gamma|a_0|^2 + i\Omega\Big)a_0 - i g\,m_0 &= 0,
\label{eq:active_ss1}\\
\Big(\frac{\gamma}{2}+i(\Delta_m-\Omega)+ iK|m_0|^2\Big)m_0 + i g\,a_0 &= 0.
\label{eq:active_ss2}
\end{align}
Equations~(\ref{eq:active_ss1})--(\ref{eq:active_ss2}) are the coupled steady-state conditions that determine $(a_0,m_0,\Omega)$. 

As the system operates under the self-oscillation condition, the photon damping is surpassed by the gain factor which can be absorbed into the \emph{effective} gain factor for simplicity (equivalent to $\kappa=0$), $G\equiv G-\kappa/2$. Further, we introduce the real auxiliary variable $A\equiv G-\Gamma|a_0|^2$, Eq.~(\ref{eq:active_ss1}) can be written compactly as
\begin{align}
(A+i\Omega)a_0 = i g\,m_0,
\label{eq:active_ss1_compact}
\end{align}
which allows one to eliminate $a_0$ in favor of $m_0$ and $(A,\Omega)$. Combining Eqs.~(\ref{eq:active_ss2}) and (\ref{eq:active_ss1_compact}) and separating real and imaginary parts yields a closed set of real constraints that, in our implementation, is reduced to a quintic polynomial equation for $A$ (see \textbf{Supplemental Material}), from which the admissible solutions and their stability are subsequently obtained.

%As the system operates under the self-oscillation condition, the photon damping is surpassed by the gain factor which can be absorbed into the \emph{effective} gain factor for simplicity (equivalent to $\kappa=0$)
%\begin{align}
%G \equiv G-\frac{\kappa}{2}.
%\label{eq:mueff_def}
%\end{align}
%Further, we introduce the real variable
%\begin{align}
%A \equiv G-\Gamma|a_0|^2 ,
%\label{eq:A_def}
%\end{align}
%which reduces the steady-state conditions to a one-dimensional polynomial problem: a quintic equation in $A$ is constructed and solved, real roots are retained within physically admissible bounds, and $(a_0,m_0,\Omega)$ are reconstructed analytically for each candidate solution.

To use the same horizontal-axis variable as in the passive case (Fig.~\ref{fig:1}), we define the active ``empty-condition photons'' as the isolated vdP steady number,
\begin{align}
n_0 \equiv \frac{G}{\Gamma},
%\qquad\text{so that}\qquad
%\mu_{\mathrm{eff}}=\Gamma n_0,
\label{eq:n0_active}
\end{align}
matching the mapping used in the phase-diagram sweeps.

%\paragraph*{Jacobian linearization and stability criterion.}
\subsection{Jacobian linearization and stability criterion}
For each steady solution of the passive and active systems, we evaluate the stability by linearizing the equations of motion in the doubled complex basis\cite{khan2018frequency}
\begin{align}
\mathbf{v}=(\delta a,\delta a^\ast,\delta m,\delta m^\ast)^{\mathsf T}.
\label{eq:doubled_basis}
\end{align}
This yields a $4\times4$ Jacobian matrix $\mathbf{J}$ (as shown in \textbf{Supplemental Material}). We compute its eigenvalues $\{\lambda_j\}$ and classify a solution as \emph{stable} when
\begin{align}
\max_j \Re(\lambda_j) < 0.
\label{eq:stability}
\end{align}
For the vdP oscillator, a neutral phase direction can produce an eigenvalue numerically closest to the origin. Consistent with the implementation, we exclude this neutral mode by discarding the eigenvalue with the smallest $|\lambda_j|$ and apply the stability criterion Eq.~(\ref{eq:stability}) to the remaining eigenvalues.

%\paragraph*{Phase-diagram construction.}
\subsection{Phase-diagram construction}
Phase diagrams in Fig.~\ref{fig:1} are generated on a two-dimensional grid in $(n_0,\Delta_m)$, with $\Delta_m$ sampled linearly and $n_0$ sampled logarithmically. The parameters used for simulating both the phase diagrams of the passive and active MP are $\kappa = 2\pi\times 1.5~\mathrm{MHz}$, $\gamma = 2\pi\times 16.5~\mathrm{MHz}$, $g=2\pi\times 30.0~\mathrm{MHz}$, and Kerr coefficient $K=2\pi\times 9.8$ nHz.
The external coupling considered for the passive system is set to $\kappa_{\rm ext}= \kappa/2$ for the critical coupling condition. The drive frequency used to convert $n_0$ to input power is $\omega_d=2\pi\times3.0~\mathrm{GHz}$. For the active system, the gain saturation term $\Gamma = 2\pi\times 2.0~\mu$Hz.  At each grid point we enumerate all admissible steady solutions and report the number of stable solutions $N_{\mathrm{s}}$ and unstable solutions $N_{\mathrm{u}}=N_{\mathrm{tot}}-N_{\mathrm{s}}$. Grid points with no admissible steady solutions are left blank. This procedure produces integer-valued maps of stable/unstable fixed-point counts for both passive and active MP models. The phase diagram in Fig.~\ref{fig:3} is constructed from the same stability analysis described above but instead of sampling $n_0$, we choose the effective gain factor $G$ as the variable for calculating the steady-state solutions with the input of the experimental parameters obtained and plotting the phase diagram for the active MP.

\section{Experimental details}\label{appendix_exp}

The half-wavelength microstrip-line cavity is fabricated on a FR-4 board with the print-circuit technique. The active component is the bipolar junction transistor (Infineon BFP842ESDH6327XTSA1) whose base-emitter voltage is fixed at 0.7 V and the collector-emitter voltage ($V_c$) is varied to control the driving power. Other components used in the devices can be found in \textbf{Supplemental Material}. The driving power $P_d = -35, -30, -25, -20, -15$ and -10 dBm correspond to $V_c = 0.173, 0.23, 0.38, 0.6, 1.3$ and 2.7 V, respectively. A 1-mm-diameter YIG sphere (Ferrisphere, Inc) with an undefined crystal orientation is used. The static magnetic field is supplied by a bench-top electromagnet (TINDUN WD-100W) and the field strength is calibrated by a commercial magnetometer (Beijing Cuihai Zhongyi Technology Co., Ltd. CH-1600). The reflection measurement of the YIG magnon mode through the port 3 and the transmission measurement of the cavity through the ports 1 and 2 are conducted using a microwave analyzer (Keysight N9917A). The emission of the active MP system is measured via the port 1 with a spectrum analyzer (Agilent E4440A). The magnon damping rate $\gamma$ is extracted by fitting the reflection spectrum ($|S_{11}|$) based on the input-output theory\cite{gardiner1985input} 
\begin{align}
    S_{11}(\omega) = 1 - \frac{\kappa_\textrm{a}}{i(\omega-\omega_{m})+\kappa_\textrm{load}/2},
\label{eq:S11}
\end{align}
where $\kappa_\textrm{load} = \kappa_\textrm{a} + \gamma$ and $\kappa_\textrm{a}$ is the coupling rate of the loop antenna.

\section{Simulation of the active MP dynamics}\label{appendix_sim}

The magnetic-field-dependent active MP dynamics is revealed by the frequency offset and spectral appearance of the vdP oscillator. To simulate the oscillator frequency shift $\Omega/2\pi$ during an upward magnetic-field sweep, we time-integrate the active MP dynamical model introduced above [Eqs.~(\ref{eq:passive_m}) and (\ref{eq:active_a})], and implement a \emph{memory-modified magnon detuning} such that the detuning used at the current field step depends on the oscillator shift obtained at the previous step. The integration is performed in the rotating frame of the bare oscillator frequency (i.e., the free-running vdP oscillation reference). In this frame, the nominal magnon detuning is defined as
\begin{align}
\Delta_m^0(B)\equiv \omega_m(B)-\omega_d^0,
\end{align}
and then we discretize the field sweep as a list of nominal detunings $\{\Delta_{m,k}^0\}$ and integrate
Eqs.~(\ref{eq:passive_m})--(\ref{eq:active_a}) step-by-step. The memory effect is implemented by replacing the detuning in the ODEs by an \emph{effective detuning}
\begin{align}
\Delta_{m,k}^{\mathrm{eff}}=\Delta_{m,k}^0-\Omega_{k-1},
\label{eq:mem_detuning}
\end{align}
where $\Omega_{k-1}$ is the oscillator frequency shift extracted from the previous step.

To simulate the continuous field sweep and capture hysteretic response, we also consider the memory effect on the MP state\cite{yao2023coherent}:
the initial condition at step $k$ is the final state from step $k-1$,
\begin{align}
\big(a_k(0),m_k(0)\big)=\big(a_{k-1}(T_{\mathrm{total}}),m_{k-1}(T_{\mathrm{total}})\big),
\end{align}
where $T_\textrm{total}$ is the total time for the integration. At each detuning step $k$, we integrate Eqs.~(\ref{eq:passive_m})--(\ref{eq:active_a}) on $t\in[0,T_{\mathrm{total}}]$
using the Runge--Kutta solver with a sampling interval $dt=1~\mathrm{ns}$, $T_{\mathrm{total}}=8~\mu\mathrm{s}$, and only the post-transient segment $t\ge T_{\mathrm{drop}}$
(with $T_{\mathrm{drop}}=3~\mu\mathrm{s}$) close to the steady-state behavior\cite{liu2019phase} is used for the oscillator frequency offset extraction. From the post-transient time trace $a(t)$, we estimate the oscillator frequency shift using the phase-slope method.
Writing $a(t)=|a(t)|e^{i\varphi(t)}$, a single-frequency oscillation satisfies $\varphi(t)\simeq -\Omega t+\varphi_0$.
We unwrap $\varphi(t)$ and perform a weighted linear regression over the last fraction of the trace, by which the fitted slope yields $\Omega$. For visualization of the magnetic-field (magnon-detuning)-dependent oscillator offset, we compute a Hann-windowed FFT of the post-transient oscillator signal $a(t)$ at each detuning step
and stack the normalized spectra into a two-dimensional matrix $S(f,\Delta_m^0)$, which is plotted as $\log_{10}S$ over a selected frequency window.

%\nocite{*}



\end{document}